\documentclass[draft,aps,showpacs,superscriptaddress]{revtex4}

\begin{document}

\title{Comment on
"$K^- d \to \pi \Sigma n$ reactions and structure of
the $\Lambda(1405)$" \\by
S. Ohnishi, Y. Ikeda, T. Hyodo, E. Hiyama, W. Weise}

\author{J. R\'{e}vai}
\affiliation{Wigner Research Center for Physics, RMI,
              H-1525 Budapest, P.O.B. 49, Hungary}

\author{N.V. Shevchenko}
\affiliation{Nuclear Physics Institute, 25068 \v{R}e\v{z}, Czech Republic
              \email{shevchenko@ujf.cas.cz} }

\date{\today}
              
\maketitle

We are astonished by the main statement of the paper~\cite{moshenniki}:

\noindent
"We report on the first results of a full three-body calculation of
the $\bar{K}NN \to \pi Y N$ amplitude for the $K^- d \to \pi\Sigma n$
reaction, and examine how the $\Lambda(1405)$ resonance manifests
itself in the neutron energy distributions of the $K^- d \to \pi\Sigma n$
reactions."

A calculation of the same reaction within the same theoretical framework
and with the same goal was reported at several international conferences
and published in~\cite{revai} without being mentioned in~\cite{moshenniki}.

The $\bar{K} N$ potentials applied in the two calculations differ from
each other, the quality of their reproduction of experimental data favors
those of~\cite{revai}. In addition, a doubtful approximation was used
in the potentials of some of the authors of~\cite{moshenniki} in their
previous papers, which, as it seems, were used in~\cite{moshenniki} too.
A discussion on this issue can be found in~\cite{ourKNNII}.

The original parts of~\cite{moshenniki} in comparison with~\cite{revai}
are higher partial waves and an energy-dependent potential. From this point
of view~\cite{moshenniki} seems to be a natural extension of the work
done in~\cite{revai}. However the use of nonrelativistic AGS equations
for initial kaon momentum 1 GeV/c is highly questionable. Therefore,
the realization of the author's promise to take relativistic corrections
into account should have preceded the publication.

\end{document}